\documentclass[twocolumn,showpacs,aps,prl,floatfix]{revtex4}
\usepackage{graphicx}
\usepackage{subfigure}
\usepackage{epsfig}
\usepackage{psfrag}

\makeatletter

\usepackage{verbatim}

\makeatletter

\input epsf

\def\be{\begin{equation}}
\def\ee{\end{equation}}
\def\ba{\begin{eqnarray}}
\def\ea{\end{eqnarray}}

\makeatother

\begin{document}

\title{Magnetic Excitations of Stripes Near a Quantum Critical Point}

\author{D.~X.~Yao$^1$, E.~W.~Carlson$^2$, and D.~K.~Campbell$^1$}

\affiliation{
(1) Department of Physics and Department of Electrical and Computer Engineering,
Boston University, Boston, Massachusetts 02215, USA \\
(2) Department of Physics, Purdue University, West Lafayette, Indiana  47907,
USA 
}

\date{July 8, 2006}

\begin{abstract}
We calculate the dynamical spin structure factor of spin waves for
weakly coupled stripes.  At low energy, the spin wave cone intensity
is strongly peaked on the inner branches.  As energy is increased,
there is a saddlepoint  followed by a square-shaped continuum rotated
$45^o$ from the low energy peaks. This is reminiscent of recent high
energy neutron scattering data on the cuprates.  The similarity at
high energy between this semiclassical treatment and quantum
fluctuations in spin ladders may be attributed to the proximity of a
quantum critical point with a small critical exponent $\eta$.
\end{abstract}
\pacs{74.72.-h, 75.10.Jm, 75.30.Ds, 76.50.+g}
\maketitle
 
\begin{figure}[Htb]
\psfrag{Jb}{\LARGE $J_b$}
\psfrag{Ja}{\LARGE $J_a$}
{\centering
  \subfigure
  {\resizebox*{!}{0.44\columnwidth}{\LARGE{(a)} \includegraphics{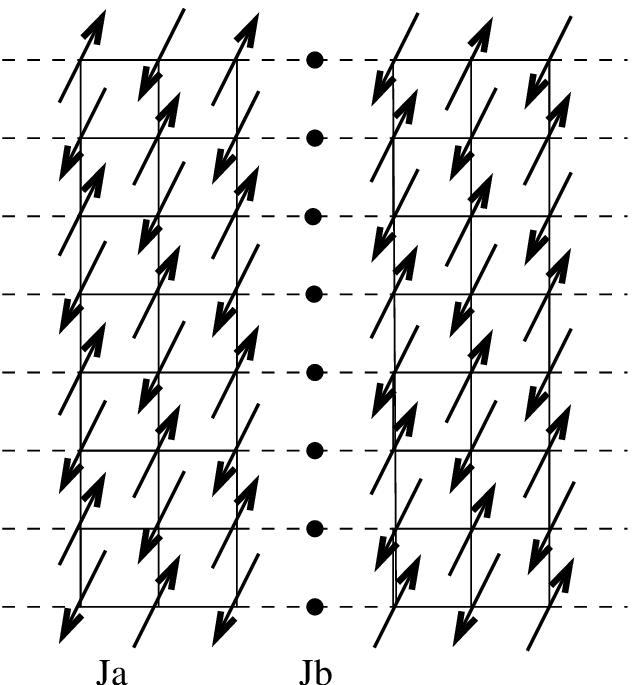}\label{fig:site}}}
  \subfigure
  {\resizebox*{!}{0.44\columnwidth}{\LARGE{(b)}\includegraphics{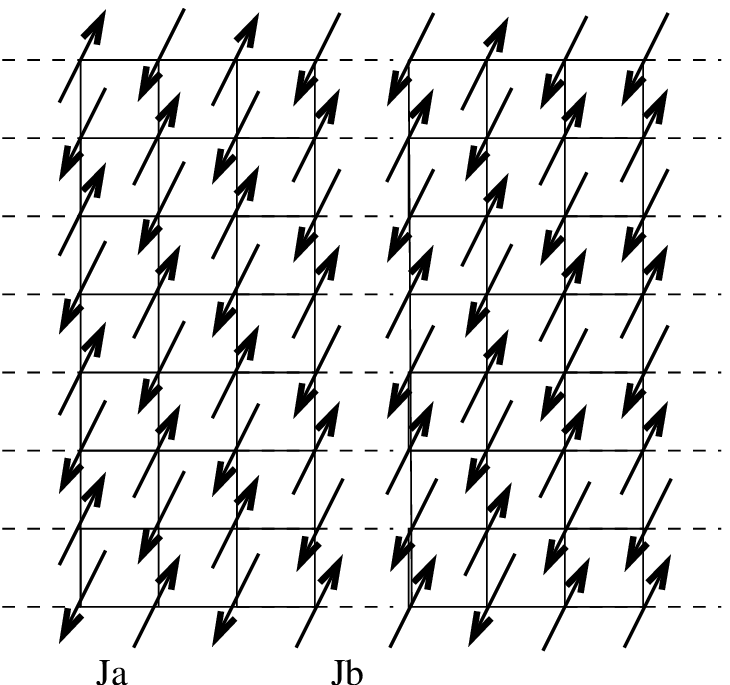}\label{fig:bond}}} 
  \par}
\caption{(a) Site-centered stripes of spacing $4$ viewed as weakly coupled 3-leg ladders. 
 (b)  Bond-centered stripes of spacing $4$ viewed as weakly coupled 4-leg ladders.  The coupling within each ladder is $J_a$, and the coupling between ladders is $J_b$.}
\label{fig:stripes}
\end{figure}

Strongly correlated electronic systems often exhibit some evidence of local electronic inhomogeneity,  
appearing in such diverse probes as neutron scattering, STM, $\mu$sR, and NMR,
among others.\cite{tranquada04a,hayden04,stripesreview}
In the high temperature superconductors, stripe structures are one possible microscopic 
realization of local electronic inhomogeneity.  Stripes are electronic states which spontaneously break
the rotational symmetry of the host crystal, the most ordered example being an interleaved,
unidirectional modulation of both spin and charge density, as shown in Fig.~\ref{fig:stripes}.
Recent neutron scattering experiments exploring the spin excitations in cuprates
point to a universal high energy magnetic response in both YBCO and (stripe-ordered)
LBCO\cite{hayden04,tranquada04a,buyers04}, and it has been suggested that the
universal high energy behavior may be connected to the superconductivity in these materials.\cite{hayden04,tranquada04a}
At low frequency, neutron scattering reveals four incommensurate spin peaks
which disperse inward toward the $(\pi,\pi)$ point, converging at a resonance peak at intermediate energy. 
This is followed by a high energy square shaped continuum in which the corners are
rotated $45^o$ away from the direction of the low energy peaks.\cite{hayden04,tranquada04a}  
The high energy response in particular has been attributed to the quantum excitations 
of spin ladders.\cite{vojta04b,uhrig04a,tranquada04a,seibold04} 

We find that the experimental results in LBCO at all energies
are  also consistent with semiclassical spin wave excitations
of weakly coupled stripes, {\em i.e.} 
for weak spin coupling across the charged domain walls in the spin pattern.  
We further suggest that the reason the quantum excitations of 2-leg ladders and 
the semiclassical spin waves studied here 
have such similar behavior at high energy  is due to the proximity of a quantum critical point (QCP) with small critical exponent $\eta$. On both the ordered and disordered side of a QCP, above a certain characteristic frequency the response is quantum critical, but for small $\eta$, the quantum critical behavior can look very much like the Goldstone behavior on the ordered side.\cite{stripefluct} Calculations of quantum spin fluctuations due to high energy ladder behavior\cite{vojta04b,uhrig04a,seibold04}
are exploring this quantum critical regime at high energy, and can resemble much of the semiclassical behavior reported here.  

In this Letter, we consider fully ordered static spin stripes, which are  
arrays of antiphase domain walls in an otherwise
antiferromagnetic texture. We are interested solely in the response
of the spin degrees of freedom and neglect the dynamics of the charge density
which must peak on every domain wall.  These charge degrees of freedom affect the spin
degrees of freedom through a spatially modulated effective exchange integral.  
The undoped system was shown to be {\em quantitatively} well described by the semiclassical 
spin waves of a $2D$ antiferromagnet.\cite{chnprl}     
When stripes form, the doping is topological, that is, doped holes create 
line dislocations in the antiferromagnetism, but 
in between the defects the antiferromagnet is more or less intact.  
Within this framework,
it is reasonable to expect a semiclassical treatment to be applicable in a range of doping,
as long as the ground state remains ordered.  
We take the coupling within
an antiferromagnetic patch to be close to its full undoped value, and consider the effective spin coupling
across a charged domain wall to be reduced, as illustrated by the dotted lines
in Fig.~\ref{fig:stripes}.
This corresponds to a Heisenberg model with modulated exchange integral on a square lattice, where
each site  represents a copper atom in the copper-oxygen plane:
\begin{equation}
H= \frac{1}{2} \sum_{\left< \mathbf{r},\mathbf{r'}\right>} J_{\mathbf{r},\mathbf{r'}} (\mathbf{S}_{\mathbf{r}} \cdot \mathbf{S}_{\mathbf{r'}})
\label{model}
\end{equation}
where \(J_{\mathbf{r},\mathbf{r'}}\) is the exchange coupling.  Nearest neighbor couplings
are positive  $J_{\mathbf{r},\mathbf{r'}}=J_a>0$ within each antiferromagnetic
patch.   Couplings across domain walls depend on whether stripes are site- or bond-centered,
as explained below.  We work in units where $\hbar = 1$.

``Site-centered'' stripes have domain walls which are centered 
on the sites of the square lattice ({\em i.e.} on the copper sites), leading to an antiferromagnetic effective coupling
$J_{\mathbf{r},\mathbf{r'}}=J_b>0$ across the domain walls.  
For weak coupling across the domain walls $J_b \ll J_a$, the system is 
close to the regime of coupled $3$-leg ladders, as shown in Fig.~\ref{fig:site}.   
``Bond-centered" stripes have domain walls which are centered between the 
sites, leading to a {\em ferromagnetic} effective coupling
$J_{\mathbf{r},\mathbf{r'}}=J_b<0$ across the domain walls.\cite{sitecentered}
For weak coupling across the domain walls $|J_b| \ll J_a$,
the system is close to coupled $4$-leg ladders,\cite{foot1}
as shown in Fig.~\ref{fig:bond}.   
When $J_b$ is sufficiently large, both cases presumably have a long-range ordered 
ground state\cite{chak-dimxover} with low energy Goldstone behavior,
but there is a quantum critical point at small $J_b$ beyond which the physics scales to that of decoupled ladders.  
While the ground state of coupled odd-leg $s=1/2$ ladders is ordered with any finite coupling and therefore  $J_b^{\rm crit} =0$,  
coupled even-leg ladders have a finite value for the quantum critical point,
$|J_b^{\rm crit}| >0$.  

We use semiclassical linearized spin wave theory and 
Holstein-Primakoff bosons, a standard procedure described
elsewhere\cite{erica04,assa,kruger03},
in order to calculate  the spin wave excitation spectrum and the 
zero-temperature dynamical structure factor, 
\begin{equation}
S(\mathbf{k}, \omega)=\sum_f \sum_{i=x,y,z} |\left<f|S^i (\mathbf{k})|0\right>|^2 \delta (\omega-\omega_f)
\end{equation}
which is related to the
expected neutron scattering intensity.  

\begin{figure}
\resizebox*{1\columnwidth}{!}{\includegraphics{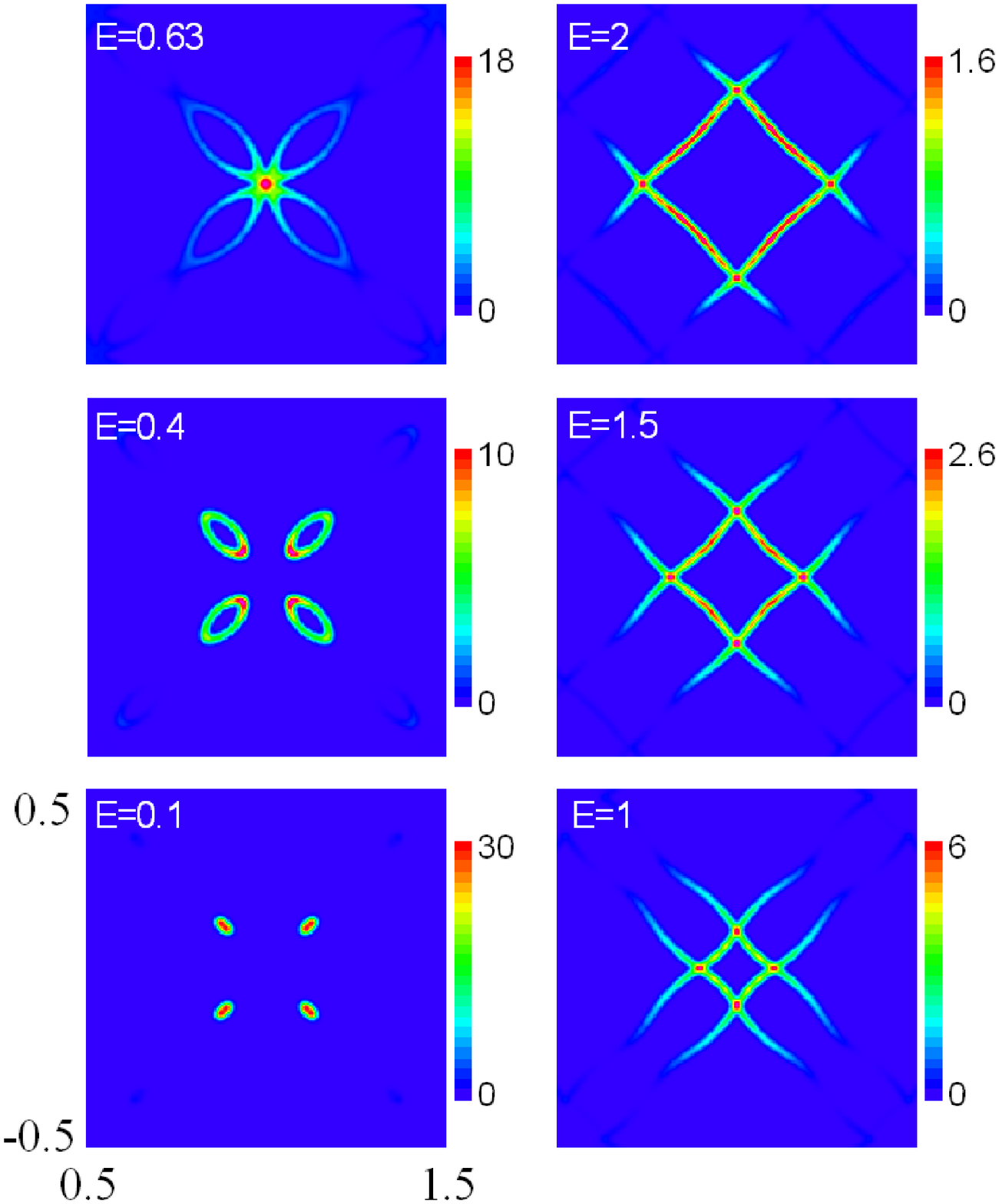}}
\caption{(Color online) Constant energy cuts, integrated within an energy window of  $\pm 0.05J_a S$ for
vertical, site-centered stripes of spacing $4$ at $J_b = 0.05 J_a$ in the
magnetic Brillouin zone.
The energy $E$ is in units of $J_a S$.  
Results are shown for twinned stripes.
\label{cut.vs4}}
\end{figure}

\begin{figure}
\resizebox*{1\columnwidth}{!}{\includegraphics{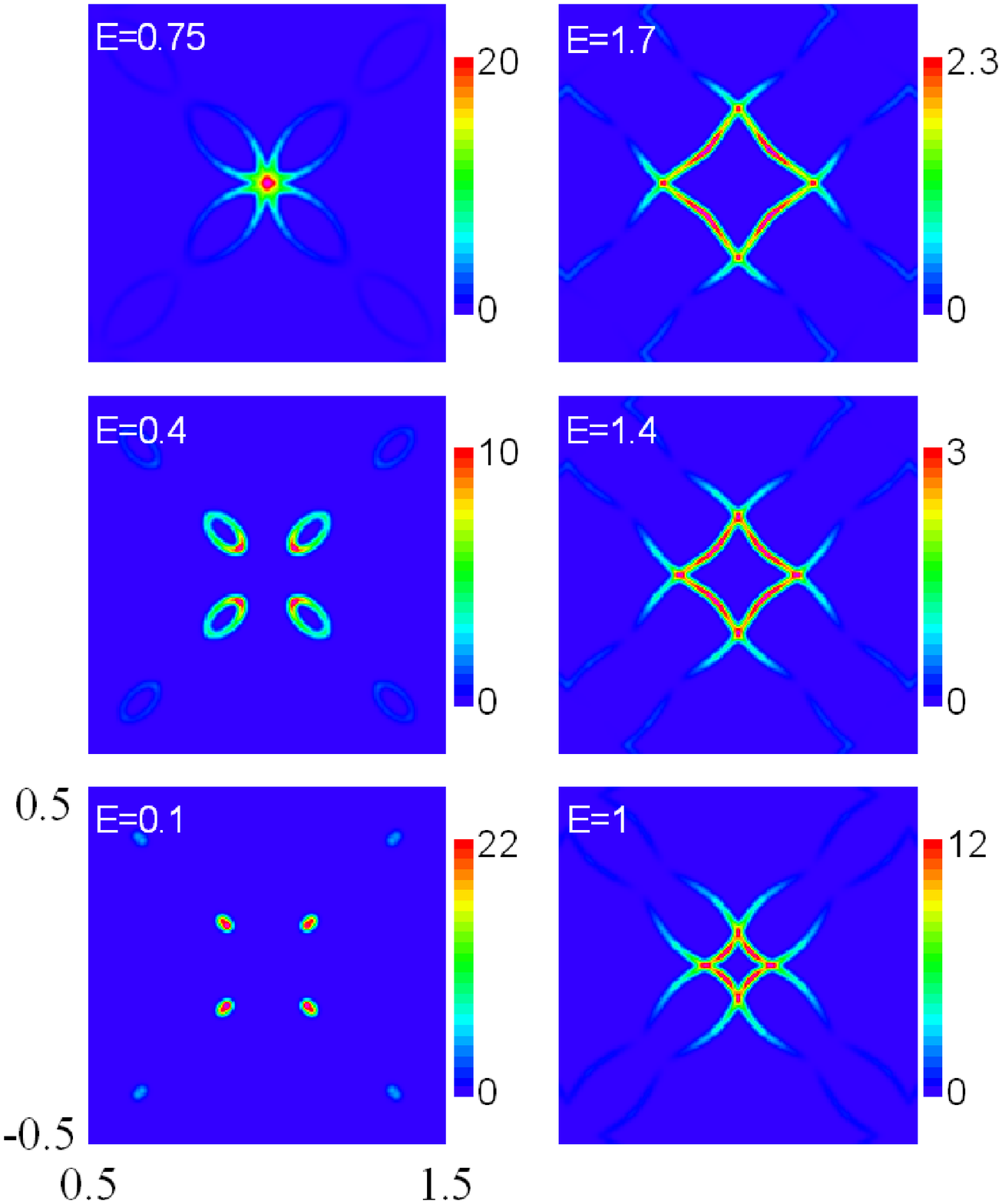}}
\caption{(Color online) Constant energy cuts, integrated within an energy window of $\pm 0.05J_a S$ for vertical,
bond-centered stripes of spacing $4$ at $J_b=-0.09J_a$ in the
magnetic Brillouin zone.
The energy $E$ is in units of $J_a S$.
Results are shown for twinned stripes.
\label{cut.vb4}}
\end{figure}

\begin{figure}
\psfrag{60}{\large $60$}
\psfrag{50}{\large $50$}
\psfrag{40}{\large $40$}
\psfrag{30}{\large $30$}
\psfrag{20}{\large $20$}
\psfrag{10}{\large $10$}
\psfrag{0}{\large $0$}
\psfrag{0.5}{\large $0.5$}
\psfrag{1}{\large $1$}
\psfrag{1.5}{\large $1.5$}
\psfrag{2}{\large $2$}
\psfrag{2.5}{\large $2.5$}
\psfrag{3}{\large $3$}
\psfrag{S}{\large $S$}
\psfrag{w}{\large $\omega$}
\resizebox*{1\columnwidth}{!}{\includegraphics{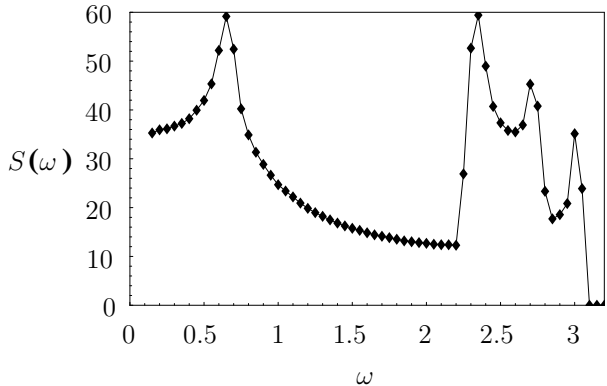}}
\caption{$S(\omega)$  for  site-centered stripes of
  spacing $4$ at $J_b=0.05J_a$ over the
magnetic Brillouin zone. The frequency $\omega$ is in units of $J_a S$.
The broadening $\Delta \omega$ is of $\pm 0.05 J_a S$.
\label{sw.vs4}}
\end{figure}

Fig.~\ref{cut.vs4} shows the intensity of the dynamical structure factor
for the site-centered stripes of Fig.~\ref{fig:site}
as a function of frequency
for weak coupling strength $J_b = 0.05 J_a$.  
Fig.~\ref{cut.vb4} shows the results for the bond-centered stripes of Fig.~\ref{fig:bond}
for weak coupling strength $J_b = -0.09 J_a$.  Both coupling strengths
have ordered ground states, as discussed below.  
The results are rather similar for site- and bond-centered stripes,\cite{vojta04b}
with the main difference being that while the satellite peaks are too weak to be visible at 
low energy in the site-centered case, they are  visible (although faint) at low energy in the 
bond-centered case.  
At low energies, four strong incommensurate peaks are visible,
which disperse inward toward $(\pi,\pi)$ as energy is increased. 
Although a spin wave cone must emanate from each magnetic reciprocal lattice vector 
due to Goldstone's theorem, the intensity is not necessarily uniform.  
For weak spin coupling across the charged domain wall,
$|J_b|  \ll  |J_a|$, we find that the intensity is strongly peaked on the inner branches
emanating from $(\pi,\pi \pm \pi/4)$, and the twinned IC peaks at $(\pi \pm \pi/4,\pi)$, 
that is, the side of the cones that is nearest $(\pi,\pi)$.\cite{future}
This situation is reversed for strong $|J_b| \gg |J_a|$, where
the intensity in the spin wave cones is strongest on the outer branches emanating
from the same points.  
One mystery about the low energy neutron scattering results in cuprates has been that spin wave
cones are not observed, but rather, the intensity disperses toward $(\pi,\pi)$.
We have shown here that this is consistent with semiclassical spin waves of coupled stripes.

Upon increasing the energy, there is a resonance peak apparent 
at $E_{\rm res} = 0.63 J_a S$ and $(\pi,\pi)$ in Fig.~\ref{cut.vs4}, 
which is a saddlepoint at which the 
integrated intensity $S(\omega)$ has a maximum (see Fig.~\ref{sw.vs4}).  
In Fig.~\ref{cut.vb4}, the resonance is at $E_{\rm res} = 0.75 J_a S$.  
The energy scale of this resonance increases with increasing $|J_b/J_a|$.
Using $E_{\rm res}= 50-60{\rm meV}$ 
from Ref.~\cite{tranquada04a}, this yields $J_a=160-190{\rm meV}$ in Fig.~\ref{cut.vs4}
and $J_a=130-160{\rm meV}$ in Fig.~\ref{cut.vb4}.  
This is  reasonable given a value of $J_a=140{\rm meV}$ 
in undoped La$_2$CuO$_4$\cite{tranquada04a}, and it is 
consistent with  other theoretical estimates on LBCO.\cite{vojta04b,uhrig04a}
For weak enough coupling $J_b$, the dispersion above the saddlepoint is highly
anisotropic, giving rise to a rotated square-shaped continuum above the {resonance~peak},
much like what is seen in LBCO.\cite{tranquada04a}.  

Fig.~\ref{sw.vs4} shows the single magnon contribution to the momentum-integrated structure factor  
$S(\omega)$ for site-centered stripes of
Fig.~\ref{cut.vs4}, 
and reveals the effect of the saddlepoint in the acoustic band. 
The broad peak in $S(\omega)$ at $E_{\rm res} = 0.63 J_a S$ is due to the saddlepoint.
Qualitatively, the results at high energy are similar to the one-triplon calculations
based on coupled $2$-leg ladders.\cite{vojta04b,uhrig04a}                                  
Contributions from the first optical band can be seen above $E=2.2J_a S$.  

\begin{figure}[Htb]
\psfrag{wG}{\large $\omega^*_{\rm G}$}
\psfrag{wQC}{\large $\omega^*_{\rm QC}$}
\psfrag{w}{\large $\omega$}
\psfrag{Jb}{\large $|J_b/J_a|$}
\psfrag{0}{\large $0$}
\psfrag{1}{\large $1$}
\psfrag{2D Antiferromagnet}{\large $2$D Magnet}
\psfrag{Disordered}{\large Disordered}
\psfrag{Ordered}{\large Ordered}
\psfrag{Goldstone}{\large Goldstone}
\psfrag{QC}{\large QC}
{\centering
\psfrag{D}{\large $\omega^*_{\rm G}$}
\subfigure
{\resizebox*{0.44\columnwidth}{!}{\LARGE{(a)}\includegraphics{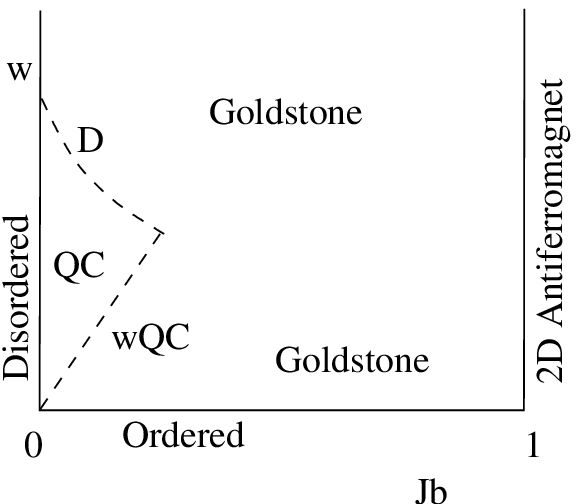}\label{fig:odd-leg}}}
\psfrag{D}{\large $\Delta$}
\subfigure
{\resizebox*{0.44\columnwidth}{!}{\LARGE{(b)}\includegraphics{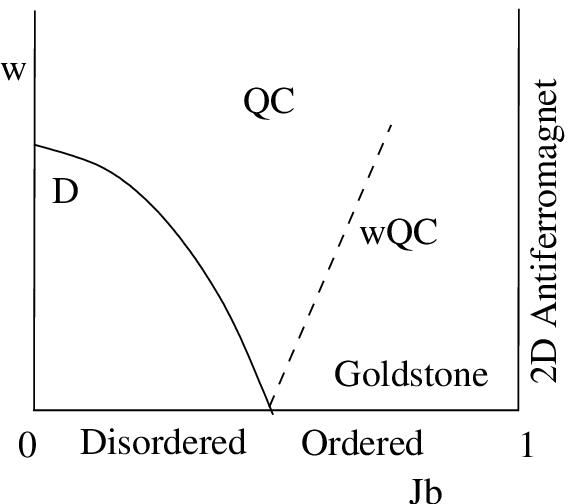}\label{fig:even-leg}}} \par}
\caption{Crossovers for coupled ladders as a function of interladder coupling $J_b$ for both odd-leg ladders  (a) and even-leg 
ladders (b).}
\label{fig:crossovers}
\end{figure}

We find that we can describe both the low energy and the high energy neutron scattering data on 
LBCO\cite{tranquada04a} within the semiclassical framework of spin waves, for weak spin coupling across the charge domain walls.  This brings up two questions:  
{1.~Why} are the semiclassical results so similar 
to the quantum critical behavior in other models of 
stripes,\cite{foot2}
and {2.~Why} is the high energy response of YBCO so similar to that of stripe-ordered LBCO, 
when YBCO does not show evidence of long range magnetic order?
Concerning the first question, the similarity may be due to the proximity to a quantum critical point
with small critical exponent $\eta$. \cite{stripefluct}
The elementary excitations of the system fundamentally change in character across the quantum critical point, 
and for $\eta > 0$ the structure factors
in the ordered and quantum critical regimes
have analytically distinct forms.     
However, the spectrum is distributed in a similar way for $\eta \gtrsim 0$.  In fact, as $\eta \rightarrow 0$, the two cases become indistinguishable. 
We are considering a 2D quantum transition to distinct 3-leg or 4-leg
ladders, and the universality class is  the
classical $(2+1)$-dimensional Heisenberg model. 
Since then the critical exponent $\eta =
0.037$  is small,\cite{sandvik03a} for all practical purposes there may be little distinction between the classical  and quantum critical cases.
However, sufficient resolution in lineshapes can in principle distinguish:  whereas weakly interacting spin waves produce a Lorentzian lineshape, quantum criticality produces a power law cusp.\cite{stripefluct,sachdev00a,sitecentered}
Concerning the second question, the gap in YBCO indicates that it must be on the
disordered side of the QCP.  However, the fact that YBCO's high energy response is so similar to the Goldstone modes calculated here is likely an indication that YBCO is  close to the QCP.

Fig.~\ref{fig:odd-leg} shows the crossover energy scales  for coupled odd-leg ladders as a function of  
interladder coupling $J_b$.   
In this case, the quantum critical point is at $J_b^{\rm crit} = 0$, where the system breaks up into independent odd-leg ladders for $S=1/2$.  At $J_b = 0$, the low frequency behavior is controlled by the $c=1$ Wess-Zumino-Witten (WZW)  model, and the high frequency behavior scales to that of weakly interacting Goldstone modes.\cite{chak-dimxover}  Away from the QCP, 
the ground state is ordered and the lowest energy response is due to Goldstone modes.  
Quantum critical behavior obtains above a lower crossover $\omega^*_{\rm QC}
\propto (J_b)^{\nu z}$,\cite{sachdev00a} 
and the high energy Goldstone behavior of decoupled odd-leg ladders is recovered above a higher crossover energy scale 
$\omega^*_{\rm G} \propto {v \over \xi} \propto v e^{- W_{eff}}$,\cite{chak-dimxover}
where the effective width of ladders $W_{eff}$ increases as $J_b$ increases.  
The two energy scales $\omega^*_{\rm QC}$ and $\omega^*_{\rm G}$
approach each other with increasing $J_b/J_a$, so that beyond a certain point 
the WZW behavior is squeezed out entirely,\cite{chak-dimxover} and in this region our approach is well-justified. 
Moreover, because the critical exponent $\eta = 0.037$ is small, even the quantum critical WZW region can resemble the Goldstone behavior.\cite{stripefluct}

Fig.~\ref{fig:even-leg} shows the gap $\Delta$ and the quantum critical energy scale $\omega^*_{\rm QC}$
for coupled even-leg ladders as a function of  $J_b$.\cite{rice-ladders}   
In this case, the quantum phase transition is at finite coupling $J_b^{\rm crit}>0$.
To find the quantum critical point for coupled $4$-leg ladders,  we use the stochastic series  expansion quantum Monte Carlo (QMC) method.\cite{sandvikqmc}
Using  finite size scaling on the spin stiffness and the Binder ratio,  we find
that $J_b^{\rm crit} = 0.076(3) J_a$
for weakly coupled $4$-leg spin-1/2 ladders, consistent with Ref.~\cite{leeqmc}.
(For weakly coupled $2$-leg ladders, $J_b^{\rm crit} \approx 0.25 J_a - 0.3 J_a$.\cite{zaanen99, rice-ladders})
Our spin-wave calculations presented in Fig.~\ref{cut.vb4}
correspond to weakly-coupled $4$-leg ladders, close to but on the ordered
side of the QCP. 
Using QMC on this configuration for systems up to size $64\times 32$, 
we find that the sublattice magnetization is $m_z = 0.143$
($47\%$ of $0.307$ for the 2D antiferromagnet), confirming that the  
system is ordered at this coupling.
 
Quenched disorder in the form of dopant atoms  introduces disorder in the ladder widths, positions, and couplings.  
Weak disorder is irrelevant at the critical point of a $(2+1)$-dimensional
Heisenberg model, and although disorder can lower the transition temperature 
(a finite ordering temperature is possible with any weak $c$-axis coupling\cite{chnprl}),
the critical exponent $\eta$ is unchanged.  
Doping may even induce a QCP through the introduction of disorder, doping the ladders,
or by inducing dynamical stripes with fluctuating charge.  



In conclusion, we have studied the magnetic excitations of ordered stripes in the proximity of a 
quantum critical point.   For weak spin coupling across the charged domain walls, the low energy spin wave cones have weight which disperses inward toward $(\pi,\pi)$, in agreement with experiment.
At higher energies, the semiclassical
excitations develop saddlepoints with steep dispersions.  
The saddlepoint acts like a ``resonance peak" in that there is extra intensity due to the saddlepoint, and also due to stripe twinning.  
(However, our model does not address the observed increase 
in intensity of the resonance peak as superconductivity onsets.) 
At energies above the saddlepoint,
there is a slowly dispersing square-shaped continuum.  The proximity to a quantum critical point (either to decoupled 3-leg ladders or to weakly-coupled 4-leg ladders) with small
critical exponent $\eta$ implies that quantum critical behavior 
(and therefore quantum ladder calculations at high energy) 
can strongly resemble the semiclassical calculations presented here.

It is a  pleasure to thank  A.~Sandvik, S. Kivelson, and J.~Tranquada for helpful discussions. 
This work was supported by Boston University (D.X.Y. and D.K.C.), and by the Purdue Research Foundation (E.W.C.). 


\begin{thebibliography}{10}

\bibitem{tranquada04a}
J.~M. Tranquada, {\it et~al.\/}, {\it Nature\/}, {\bf 429}, 534 (2004).

\bibitem{hayden04}
S.~M. Hayden, {\it et~al.\/}, {\it Nature\/}, {\bf 429}, 531 (2004).

\bibitem{stripesreview}
For a review, see, {\em e.g.}, E.~W.~Carlson {\it et al.}, in The Physics of Superconductors, Vol. II, ed. J.~Ketterson and
  K.~Benneman, Springer-Verlag, 2004, and references therein.

\bibitem{buyers04}
C.~Stock, {\it et~al.\/}, {\it Phys. Rev. B\/}, {\bf 71}, 024522 (2004).

\bibitem{vojta04b}
M.~Vojta and T.~Ulbricht, {\it Phys. Rev. Lett.}, {\bf 93}, 127002 (2004).

\bibitem{uhrig04a}
G.~S. Uhrig, K.~P. Schmidt, and M.~Gr$\ddot{u}$ninger, {\it Phys. Rev.
  Lett.\/}, {\bf 93}, 267003 (2004).

\bibitem{seibold04}
G.~Seibold and J.~Lorenzana, {\it Phys. Rev. Lett.\/}, {\bf 94}, 107006 (2005).

\bibitem{stripefluct}
S.~A. Kivelson, {\it et~al.\/}, {\it Rev. Mod. Phys.\/}, {\bf 75}, 1201 (2003).

\bibitem{chnprl}
S.~Chakravarty, B.~I. Halperin, and D.~R.~Nelson, {\it Phys. Rev. Lett.\/}, {\bf
  60}, 1057 (1988).

\bibitem{sitecentered}
J.~M. Tranquada {\it et al.}, {\it Phys.
  Rev. B\/}, {\bf 55}, R6113 (1997).

\bibitem{foot1}
Coupled $2$-leg ladders have been considered in Refs.~\cite{vojta04b,uhrig04a}.
  We are considering here an effective theory of the behavior of the net moment
  on each site, with some smooth envelope for the moments which must have a
  node at each domain wall. This produces a small but finite moment even on
  sites right next to the domain walls, and so we model bond-centered stripes
  as weakly coupled $4$-leg ladders.

\bibitem{chak-dimxover}
S.~Chakravarty, {\it Phys. Rev. Lett.\/}, {\bf 77}, 4446 (1996).

\bibitem{erica04}
E.~W. Carlson, D.~X. Yao, and D.~K. Campbell, {\it Phys. Rev. B\/}, {\bf 70},
  064505 (2004).

\bibitem{assa}
A.~Auerbach, {\it Phys. Rev. Lett.\/}, {\bf 72}, 2931 (1994).

\bibitem{kruger03}
F.~Kr$\ddot{u}$ger and S.~Scheidl, {\it Phys. Rev. B\/}, {\bf 67}, 134512 (2003).

\bibitem{future}
D.~X. Yao, E.~W. Carlson, and D.~K. Campbell, {\it Phys. Rev. B\/}, {\bf 73}, 224525 (2006).

\bibitem{foot2}
Models pointing to quantum behavior at high energy include weakly coupled 
$2$-leg ladders,\cite{vojta04b,uhrig04a,tranquada04a} a phenomenological quantum lattice model,\cite{vojta-sachdev}
and Gutzwiller projections of the Hubbard model.\cite{seibold04}


\bibitem{vojta-sachdev}
M.~Vojta and S.~Sachdev, {\it J. Phys. Chem. Solids}, {\bf 67}, 11 (2006).


\bibitem{sandvik03a}
P.~Sengupta, A.~W. Sandvik, and R.~R.~P. Singh, {\it Phys. Rev. B\/}, {\bf 68},
  094423 (2003).

\bibitem{sachdev00a}
S.~Sachdev, {\it Science\/}, {\bf 288}, 475 (2000).

\bibitem{rice-ladders}
S.~Gopalan, T.~M. Rice, and M.~Sigrist, {\it Phys. Rev. B\/}, {\bf 49}, 8901
  (1994).

\bibitem{sandvikqmc}
A.~W. Sandvik and J.~Kurkij$\ddot{a}$rvi, {\it Phys. Rev. B\/}, {\bf 43}, 5950 (1991).

\bibitem{leeqmc}
Y.~J. Kim, {\it et~al.\/}, {\it Phys. Rev. B\/}, {\bf 60}, 3294 (1999).

\bibitem{zaanen99}
J.~Tworzydlo {\it et al.},  {\it Phys. Rev.
  B\/}, {\bf 59}, 115 (1999).


\end{thebibliography}
\bibliographystyle{forprl}

\end{document}